\documentclass{article}

\usepackage{arxiv}

\usepackage[utf8]{inputenc} 
\usepackage[T1]{fontenc}    
\usepackage[colorlinks=true, allcolors=blue]{hyperref}       
\usepackage{url}            
\usepackage{booktabs}       
\usepackage{amsfonts}       
\usepackage{nicefrac}       
\usepackage{microtype}      
\usepackage{lipsum}		
\usepackage{graphicx}
\usepackage{natbib}
\usepackage{doi}

\usepackage{amsmath}

\usepackage{todonotes}

\title{
Masking Behaviors in Epidemiological Networks with Cognitively-plausible Reinforcement Learning} 

\author{
  {Konstantinos Mitsopoulos}\\
  Institute For Human and Machine Cognition\\
  Pensacola, FL 32505\\
  \texttt{kmitsopoulos@ihmc.org}\\
  \And
  {Lawrence Baker}\\
  RAND Corporation, \\Boston, MA 02116\\
  \texttt{lbaker@rand.org}
  \And
  {Christian Lebiere}\\
  Carnegie Mellon University\\
  Pittsburgh, PA 15213\\
  \texttt{cl@cmu.edu}\\
  \And
  {Peter Pirolli}\\
  Institute For Human and Machine Cognition\\
  Pensacola, FL 32505\\
  \texttt{ppirolli@ihmc.org}\\
  \And
  {Mark Orr}\\
  Biocomplexity Institute\\
  University of Virginia\\
  Charlottesville, VA 22904\\
  \texttt{mo6xj@virginia.edu}
  \And
  {Raffaele Vardavas}\\
  RAND Corporation, \\Santa Monica, CA 90401\\
  \texttt{rvardava@rand.org}\\ 
  \And
}

\date{}

\renewcommand{\shorttitle}{Masking Behaviors in Epidemiological Networks with Cognitively-plausible Reinforcement Learning}

\begin{document}
\maketitle

\begin{abstract}
The COVID-19 pandemic highlighted the critical role of human behavior in influencing infectious disease transmission and the need for models capturing this complex dynamic. We present an agent-based model integrating an epidemiological simulation of disease spread with a cognitive architecture driving individual mask-wearing decisions. Agents decide whether to mask based on a utility function weighting factors like peer conformity, personal risk tolerance, and mask-wearing discomfort. By conducting experiments systematically varying behavioral model parameters and social network structures, we demonstrate how adaptive decision-making interacts with network connectivity patterns to impact population-level infection outcomes. The model provides a flexible computational framework for gaining insights into how behavioral interventions like mask mandates may differentially influence disease spread across communities with diverse social structures. Findings highlight the importance of integrating realistic human decision processes in epidemiological models to inform policy decisions during public health crises. 
\end{abstract}

\section{Introduction}

Recent advances in data science have enabled new ways of studying social systems, from population-based networks to information spread on social media. However, while this data provides detailed snapshots, simulations are needed to systematically explore unobserved spaces of possibilities that human dynamics can evolve into. For example, the recent COVID-19 pandemic has introduced unprecedented challenges, not only in its epidemiological spread but also in the critical role of human behavioral responses to mitigate its transmission \citep{aledortNonpharmaceuticalPublicHealth2007}.
This highlighted the importance of modeling human behavior like social distancing and vaccination adherence at population scales, as individual decisions modulate pathogen transmission dynamics. More sophisticated models that capture the heterogeneity and complexity of human psychology could thus inform policymaking in creating effective strategies to combat infectious diseases~\citep{manheimImprovingDecisionSupport2016} without excluding other civically-important domains including public health, climate change, cybersecurity, and more.

Disease transmission is not solely driven by biological factors but is also determined by human behavior and social networks. Among the nonpharmaceutical interventions (NPIs) utilized, mask-wearing has emerged as a pivotal measure in slowing and reducing the virus's spread and transmission \citep{alagozEffectTimingAdherence2020}. While various models have attempted to forecast disease spread and estimate the impact of interventions \citep{manheimImprovingDecisionSupport2016}, there's a significant gap in understanding how adaptive behaviors, particularly related to mask-wearing decisions, interact with diverse network structures and influence disease epidemiology \citep{craneChangeReportedAdherence2021}.

In this study, we aim to explore the interplay of these factors using an enhanced Agent-based Modeling (ABM) approach. Specifically, we simulate epidemiological disease transmission on networks integrated with adaptive cognitive agents focused on mask-wearing decisions. The agents are grounded in a cognitive architecture with decision-making based on optimizing a utility function encoding intrinsic motivations and perceptions. Moreover, agents continually update their behavioral policies over time using Reinforcement Learning (RL), enabling adaptation to the developing dynamics of the pandemic environment based on experiential feedback. This integrated framework allows examining the interdependencies between individual-level public health decisions, network-level social influences, and population-level disease transmission outcomes.

\section{Background}
Computational epidemiology, a multidisciplinary domain, explores various facets of epidemiology, including disease spread and the effectiveness of public health interventions \citep{manheimImprovingDecisionSupport2016}. Understanding the trade-offs between policies and hypothetical scenarios requires models rooted in causal epidemiological and behavioral theory, supported by relevant data \citep{pearlCausality2009}. This necessity arises from the complex interplay of socio-psychological behavioral dynamics inherent in real-world disease transmission, encompassing a range of factors like demographics and spatial considerations. The effectiveness of interventions isn't solely dependent on epidemiological data; it's equally shaped by individual responses, prevailing social norms, and societal constraints \citep{squazzoniComputationalModelsThat2020}. To address these limitations, there is a need to combine epidemiological models of disease transmission with those describing adaptive behaviors \citep{chenChapter12Individual2017,vardavasModelingInfectiousBehaviors2021a, verelstBehaviouralChangeModels2016}. The research field of combining disease epidemiology models with models of adaptive human behavior is more than a decade long \citep{ manfredi2013modeling,funkModellingInfluenceHuman2010, bauchGroupInterestSelfinterest2003, relugaEvolvingPublicPerceptions2006, vardavasCanInfluenzaEpidemics2007}. However, now in the wake or aftermath of the COVID-19 pandemic this field is being reawakened and rediscovered and there is a recognized need to combine these epidemiological models and behavioral models to better inform policy. In particular their is a need to model how compliance levels across populations with different policy interventions changes over time and how this affects the disease epidemiology \citep{becherComparativeExperimentalEvidence2020}. Compliance varies significantly due to complex sociological dynamics influenced by demographics, peer pressure, risk perceptions, and beliefs about intervention efficacy \citep{chanWhyPeopleFailed2021}.

Two primary types of simulations used in this context are mechanistic epidemiological models: (i) compartmental population-based models (PBM) and (ii) individual-level agent-based models (ABMs). Traditional simulations in this field have leaned on PBMs, relying on differential equations to represent disease characteristics like infectivity and transmissibility \citep{adigaMathematicalModelsCOVID192020}. However, PBMs operate at an aggregate level, lacking individual-level adaptation. Although they can incorporate some population heterogeneities, they do not capture the complete complexity of real networks and population mixing and clustering. Moreover, while descriptive behavioral models can be coupled with PBMs by adaptively changing the disease transmissibility over time in response to epidemiological outcomes, they do so at the aggregate level, not allowing for individual-level adaptation \citep{nowakOptimalNonpharmaceuticalPandemic2023}.
 
More realistic simulations necessitate an individual-level framework whereby individuals interact, mix, and cluster in complex ways which can be described by realistic networks, and where individuals make autonomous decisions as agents \citep{cornforthErraticFluVaccination2011, vardavasModelingInfluenzaVaccination2013}. This has prompted the need for more sophisticated models with deliberative agents to capture human decision-making realistically, where variability in behaviors and decisions can emerge due to differences in individual epidemiological histories and trajectories instead of being accounted for by belonging to different aggregate-level population groups.
 
To address these limitations, ABMs have emerged as a pivotal approach in computational epidemiology \citep{auchinclossBriefIntroductoryGuide2015, gaudouCOMOKITModelingKit2020, hinchOpenABMCovid19AgentbasedModel2020, limaImpactMobilityRestriction2021}. Especially during the COVID-19 pandemic, the significance of ABMs in understanding the impact of behavioral health interventions has been pronounced. Consequently, a surge in ABMs has aimed to predict intervention effects by considering factors like age, household profiles, and evolving interaction patterns over time and space. However, despite this proliferation, existing models of individual decision-making remain limited.

ABM typically involves defining specific ad-hoc rules for agent interactions and observing the resultant behaviors through simulations. While effective, this approach can limit the emergence of complex and adaptable behaviors, thus constraining the exploration of the full spectrum of possibilities. Reinforcement learning (RL), a Machine Learning method where agents learn to make decisions by receiving rewards or penalties for their actions, complements ABM by by enhancing the adaptability and complexity of agent behaviors. The feedback guides the agent in learning which actions lead to the most favorable outcomes over time. RL is vital for understanding behavior as it mimics the way organisms learn from the consequences of their actions in real life). By using RL, we can create models that predict and analyze complex behaviors, offering insights into decision-making processes in diverse contexts, from gaming and robotics to social sciences and economics.

Still, conventionally realized RL agents lack integration of real individual-level psychological theory required for human-like behavior. Here we combine ABM with a cognitive architecture grounded in utility learning theory to demonstrate the importance and interplay of data-driven behavioral models embedded in realistic social networks. In particular, we simulate the adoption of mask wearing behavior during an epidemic across a population whose interactions are informed by data. Each agent chooses to mask or unmask based on a utility function integrating individual desire for group conformity, perceived infection risk, and wearing discomfort. We conduct experiments varying the population composition and explore resulting infection outcomes. This demonstrates the capabilities of situated, theory-grounded cognitive models within population-scale simulations for exploring possibility spaces of human behavior change. Findings can inform public health policy decisions amidst crises requiring mass behavior change.

 
 

Our research aims to uncover the complex dependencies between human behavioral responses regarding mask-wearing and disease spread across diverse network topologies \citep{bhattacharyyaMathematicalModelsInterplay2012}. This exploration is crucial in informing targeted interventions and policy formulations for managing infectious disease outbreaks, especially in the context of COVID-19 and future pandemics \citep{kisslerProjectingTransmissionDynamics2020}. 
Moreover, an intriguing aspect of our investigation lies in discerning potential interacting effects between adaptive behavioral models and network structures on disease epidemiology. By systematically manipulating key behavioral parameters and network morphologies, we seek to unravel the interplay between these elements, unveiling emergent phenomena that could significantly impact disease transmission dynamics.

\section{Methods}
We develop an utility learning-based model in which agents make decisions about mask wearing based on balancing competing preferences. Each agent receives ongoing inputs about the global pandemic status, like infection rates, and local status through the number of sick individuals in their area. Their decision-making is based on a utility function integrating considerations such as conforming to neighbor behaviors, discomfort from extended mask usage, and personal infection risk tolerance. By tuning these utility parameters
and situating agents in social network topologies, we promote different motivations that drive individual variation and changes in protective behaviors over time. The agents continually learn which health behaviors maximize their utilities by leveraging memories of past behavioral outcomes stored in their instance-based cognitive architecture. By conducting such simulations we can observe emerging infection dynamics at population scales under different compositions of decision motivations and social structures. This allows testing hypothesized mechanisms that drive societal-level adherence and evaluating policy options to encourage protective behaviors critical to pandemic response.

\subsection{Agent-Based Modeling}

We employ an agent-based SEIR (Susceptible, Exposed, Infectious, Recovered) epidemiological model, where the disease percolation takes places across a network. Agents in the model represent an individual who can transition through each of the SEIR states. The infectious period encompasses pre-symptomatic, symptomatic, and asymptomatic phases. This model operates on a daily timestep, enabling a granular simulation of both disease progression and percolation, the durations of these disease states are detailed in Table 1 and are geometrically distributed. These disease characteristics are not modeled on a specific disease, but are chosen to be representative of a disease which could cause a global pandemic. Individuals can be infected if they are susceptible and one of their linked neighbors is infectious. After infection, they are granted sterilizing immunity for 75 days, after which they can be reinfected.

\begin{table}[h]
\centering
\caption{Epidemiological ABM Parameters}
\label{your-table}

\begin{tabular}{ll}
\toprule
Variable                        & Value \\
\midrule
\multicolumn{2}{l}{\textbf{Disease state duration}} \\
Exposed                         & 2 days \\
Presymptomatic infectious       & 3 days \\
Symptomatic infectious          & 8 days \\
Asymptomatic infectious         & 8 days \\
Sterilizing Immunity            & 75 days \\
\midrule
\multicolumn{2}{l}{\textbf{Transition Probabilities}} \\
Asymptomatic proportion         & 0.2 \\
\midrule
\multicolumn{2}{l}{\textbf{Disease spread}} \\
Basic reproduction number       & 5 \\
Random mixing proportion        & 20\% \\
Initial Exposed proprtion       & 1\% \\
Masking effectiveness           & 80\% \\
\bottomrule
\end{tabular}

\end{table}

The network on consists of nodes representing agents and edges representing potential contacts between agents, with edge weights signifying the daily probability of transmission. The primary network in our study is a synthetic socio-centric graph developed by the Network Dynamics and Simulation Science Lab at Virginia Tech, which intricately outlines the social connections in Portland, Oregon \citep{maratheSyntheticDataProducts2014}. This dataset serves as a comprehensive representation of daily social interactions in an urban setting, encompassing everything from close-knit friendships to chance encounters. This dataset significantly enriches our understanding of how social connections form and evolve within the city and has previously been used to model infectious disease transmission dynamics \citep{eubankModellingDiseaseOutbreaks2004}.

This dataset represents the entire city of Portland, over 1.6 million individuals, which is too large to simulate in a timely manner. To streamline this extensive network to around ten thousand individuals, we employed a method involving the selection of specific clusters from the original network. These clusters encompassed all internal edges connecting the sampled nodes and the peripheral edges or stubs.We proceeded on an iterative process to connect these stubs while aiming to preserve the original dataset's network degree distributions and demographic mixing matrices. This method essentially restructures connections within the network to maintain the fundamental structure of the original dataset. It involves two primary steps: first, reorganizing connections between nearby nodes, focusing on links between the inner and outer segments of the sampled network. Careful recalibration of these connections involves using calculated edge weights to establish meaningful associations between nodes linked to external ones. The second step entails rewiring connections between distant nodes, particularly those without commonalities with external nodes. This phase involves a thorough examination of the degree structures of these nodes, attempting to adjust connections while upholding the observed degree distributions in the original dataset. This iterative process aims to minimize disparities between the original and modified networks, ensuring that the distilled network maintains essential characteristics while providing a more manageable portrayal of Portland's intricate social network. This process yielded a network with 9,223 individuals and 102,623 edges. We generated alternative networks so that we could explore the impact of network structure on both disease percolation and learning processes. This included random unweighted and Barabási-Albert Scale-Free graphs, which were sampled such that the total number of nodes and edges matched the Portland network.

We calibrated these networks to reproduce a specified basic reproduction number (R0) by uniformly scaling the weights of all edges, ensuring that we can generate realistic transmission dynamics. An edge between a susceptible and an infectious individual has the potential to be 'realized'—that is, to result in a transmission event—based on its weight. Masking is modelled as reducing the probability of both infecting others and being infected, such that two individuals who are masked are unlikely to infected one another. Social network data can struggle to capture low probability infections - such as the small chance that a single person infects each other person in a crowded public space like a concert venue or supermarket. To capture these interactions, we assign 20\% of the R0 to random mixing. For random mixing, we calculate the expected number of infections based on the R0, number of infected people, number of susceptible people, and aggregate mask wearing behavior. We then randomly assign these expected infections to susceptible individuals throughout the network. By using network and random mixing together, we can capture the complex stochastic nature of disease spread through a social network.


\subsection{Psychologically-Valid Agents based on Cognitive Architectures}
PVAs \citep{pirolli2020cognitive, pirollimining2021, pirolliPVAs2022, mitsopoulos_pvgas_2023} are computational agents implemented within the ACT-R architecture \citep{anderson2004integrated} to simulate and analyze human behaviors. They offer a modeling approach with input drivers induced from heterogeneous sources including online media, demographics, psychological traits etc. The subset of ACT-R methods employed in designing PVAs is grounded in the Instance Based Learning Theory \citep{gonzalez2003instance}. We refer to this specific approach as CogIBL to differentiate it from other Instance-based Learning methods prevalent in the Machine Learning field.

The CogIBL model is based on the idea that decisions and behaviors have subjective utility (or value), such as satisfaction or preference. When a behavior occurs in a situation and produces an outcome, it is associated with a subjective assessment of its value. Following ACT-R theory, these experiential associations are stored in \textbf{declarative memory} as experiential records (chunks) of decision-making situations, behaviors, outcomes, and their values.

Over time, this repository of experiences forms the basis for implicit and explicit knowledge about decision-making \citep{lebiere1998implicit, lebiere1999implicit, wallach2003conscious}. It is assumed that when individuals are faced with decisions, they draw from these stored experiences, retrieving memories that align with current cues to evaluate alternatives and decide on actions. This relies on ACT-R's memory \textbf{retrieval} and \textbf{blending} mechanisms. Retrieval uses situation cues to recall past instances based on their \textbf{recency}, \textbf{frequency} and \textbf{similarity} to the current situation. Blending aggregates and generalizes across activated memories. By leveraging instance-based  knowledge the model is able to estimate expectations of potential outcomes based on past similar situations.

\begin{figure}
\centering
\includegraphics[scale=1]{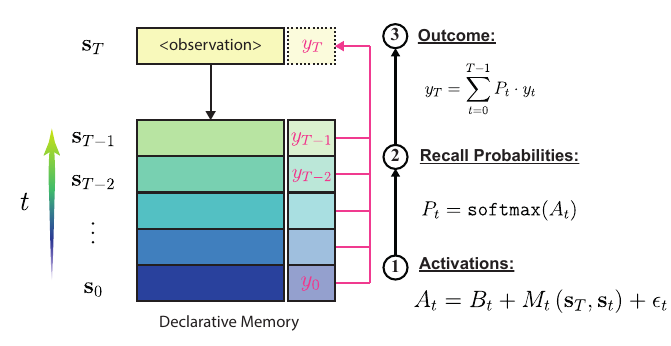}
\caption{An Overview of Instance-Based Learning Processes. IBL theory argues that implicit expertise is gained through the accumulation and recognition of previously experienced events. Events are stored in the Declarative Memory and are retrieved, weighted accordingly, in order to generate the model's response.}
\label{images:ibl}
\end{figure}

In Figure \ref{images:ibl} we describe in detail the computations that take place in the CogIBL model. In brief, CogIBL implements a Kernel Smoother, a non-parametric instance-based, Supervised Learning function approximation method. In our setting we use the model to approximate the utility for actions related to masking.
Assuming that we seek an outcome $y_T$ given a new instance $\textbf{s}_T=(f_1,...,f_k,...,f_K)$ with attributes (features) $f_k$, inference is performed in three main steps:

\begin{enumerate}
    \item \textbf{Activations Computation:} Each stored prior experience (instance) has an activation $A_t$ indicating its relevance to the current situation. This depends on two components: the \textit{base-level activation} $B_t$, a function of recency and frequency of the instance's use; and the \textit{Matching Score} $M_t(\textbf{s}_T$, measuring similarity between the current state $\textbf{s}_T$ and the stored instance state $\textbf{s}_t)$ based on a distance metric. The activation is a real-valued combination of these cues with stochastic noise $\epsilon_t$ added, modeling stochastic memory recall. In our model, we set $B_t=0$ to solely leverage the current context without biases from past instance use. Further, we remove the noise to weigh instances by matching scores rather than introducing randomness.
    
    \item \textbf{Retrieval Probabilities:} Activations are normalized by using the softmax function.  
    
    \item \textbf{Blending:} Decision output is the weighted average of past decisions $y_t$, weighted by their relevance to the current situation via recall probabilities. This outcome minimizes directly the mean squared error between model's estimation and observed output. In our case, the model estimates the action-values for masking/unmasking and takes the form:
    
    \begin{equation}
        \hat{Q}_t(\mathbf{s},a)=\sum_{t=0}^{T-1}P_t \cdot Q_t
        \label{blending}
    \end{equation}
    
\end{enumerate}

\begin{figure}[ht]
    \centering
\includegraphics[scale=0.3]{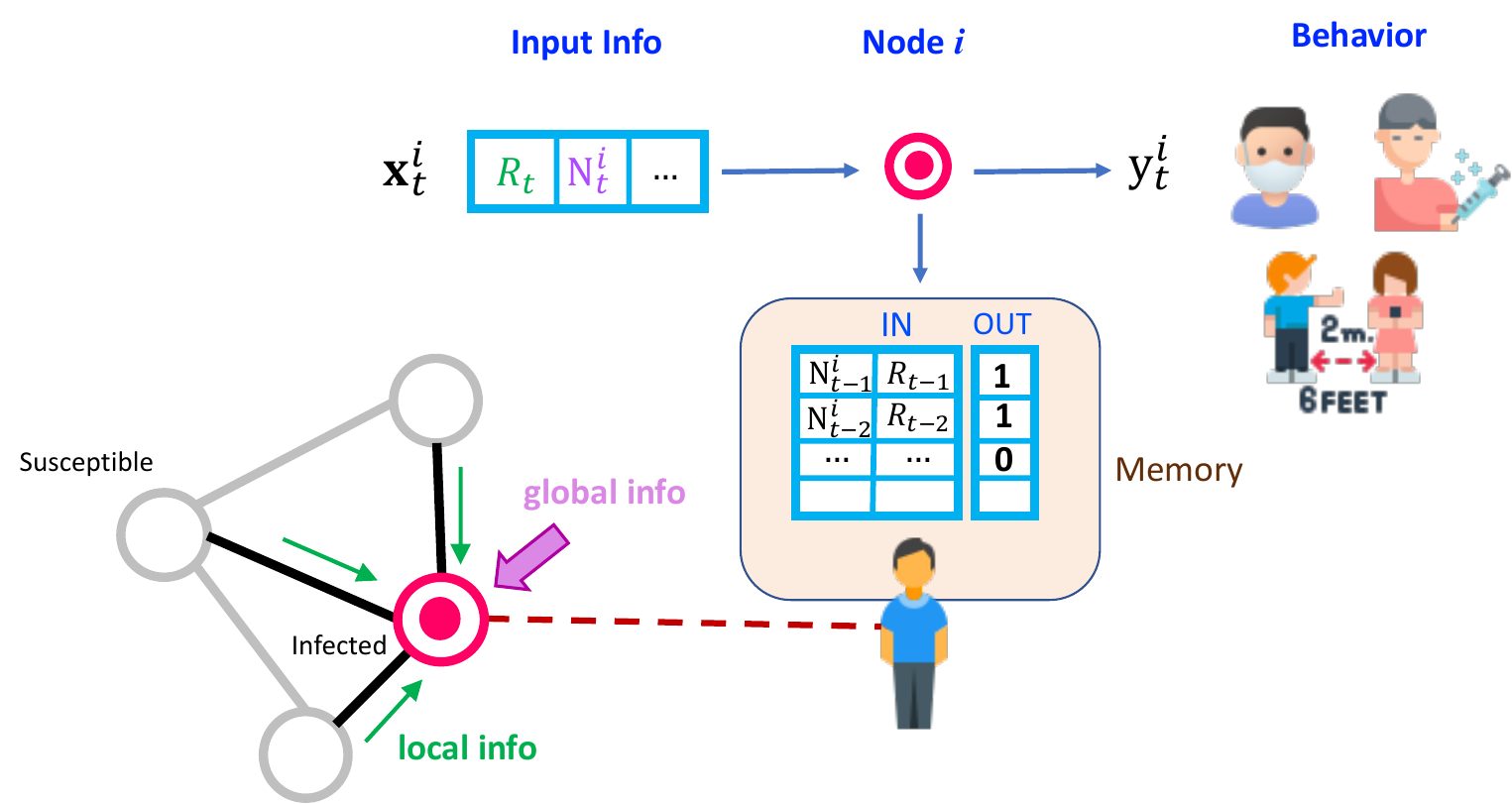}
    \caption{Example of PVA functionality in ABM simulation.}
    \label{fig:ABM_IBL}
\end{figure}

At every timestep, agents perceive the current state of the system as a vector $\mathbf{s}_t=\left(  M_{\text{local}}, I_{\text{local}}, I_{\text{global}} \right)$ of proportion of masked and infected neighbors and global proportion of infected population, combining local and global information from the disease transmission network. An example is depicted in In Figure \ref{fig:ABM_IBL}. Each agent estimates the action-value function $Q(s,a)$ that indicates how preferable is for the agent to (un)mask given the current state of the pandemic. A typical learning mechanism of an RL agent is Q-Learning \citep{watkins1992q}, which updates the Q-values using the following equation:
    
\begin{equation}
    Q(s, a) \leftarrow Q(s, a) + \alpha \left( R(s, a) + \gamma \max_a Q(s', a) - Q(s, a) \right)
    \label{qlearning}
\end{equation}
where $\alpha$ represents the learning rate, and $R(s,a)$ is the reward function. However, due to the continuous nature of our task, enumerating all possible states and actions becomes impractical. To address this challenge, we employ CogIBL's estimation capabilities to approximate the action value function. This involves formulating the problem as a regression task, where the Kernel Smoother's output (eq. \ref{blending}) minimizes the mean squared error (MSE) between received rewards and estimated rewards. 

This approach conceptually aligns with Deep Q-Learning \citep{mnih2013playing}, where action values are estimated by a parametric neural network that approximates the Q-function. However, our framework alternatively leverages the non-parametric, instance-based kernel regression native to our cognitive architecture. This enables cognitively-plausible RL within the agent-based modeling simulation while preserving interpretability of the emerging behaviors in terms of cognitive constructs. The data-driven kernel estimations mesh well with the rapidly evolving pandemic statistics requiring dynamic adaptation. Further, learning relies on comparing new experiences to the agent's memory rather than propagating gradients through layers of predefined parameters. This mirrors human-like rapid decision adjustment based on accrued observations.

Kernel smoothers are non-parametric, instance-based learning models. This means that, in contrast to parametric models, they do not undergo a distinct training phase. Instead, they require accumulating examples in a memory repository to subsequently leverage for estimations. For this reason, we pre-populate all agents' memories with the true utility values for the extreme cases (boundaries) of each state variable. As the state variables in our model represent proportions bounded between 0 and 1, the edge cases correspond to values of 0 and 1 for each variable. Seeding these boundary utility assessments ensures agents can interpolate to novel intermediate states based on available memories, without requiring a prolonged independent training period.

\subsection{Decision Making}
Our hypothesis is that agents do not extensively plan for the longer-term future when deciding whether to wear a mask. Instead, they assess criteria relevant to the present moment, based on the local and global pandemic information they receive. To capture this short-term reward optimization, we set the discount factor $\gamma=0$ and make rewards dependent solely on the immediate state rather than future states. In contrast to account for cumulative reward after multiple decision steps we set $\gamma=0.95$ and adjust the reward function to account for a complete state-action-next state transition.

Each agent follows a certain policy given by:

\begin{equation}
    P(a|s) = \frac{e^{\beta Q(s, a)}}{\sum_{a'} e^{\beta Q(s, a')}}
    \label{policy_boltz}
\end{equation}

where $\beta$ is the exploration-exploitation trade-off parameter. For our purposes it was set to $\beta=5$ so the agents are leaning towards exploitation.



\subsection{Reward Function}
At every step, the agents receive a scalar reward value as feedback for their action. We assume that mask-wearing is a behavior that depends on a multitude of factors which have to do with the internal reward system of each individual rather than external factors. For this, we define an intrinsic reward function that we provide to agents based on evaluating their current state and actions regarding mask-wearing decisions. This scalar utility results from the weighted sum of three key reward components:
\begin{equation}
   R(s,a) = -w_1 \cdot \text{DP} + w_2 \cdot \text{CR} + w_3 \cdot \text{RR} 
\end{equation}

The reward components are deifned as follows:
\begin{itemize}    
    \item \textbf{Discomfort penalty (DP):} This penalty represents the relative agent’s discomfort with mask-wearing. DP is defined as $\text{DP} = -a$

    \item \textbf{Conformity reward (CR):} This reward promotes an agent’s conformity to the mask-wearing behaviors of neighboring agents. CR is defined as $\text{CR} = 1 - \left| a - M_{\text{local}} \right|$ where $M_{\text{local}}$ is the proportion of masked neighbors.
    
    \item \textbf{Risk reduction reward (RR):} This reward promotes an agent’s perception of infection risk reduction from wearing masks. RR is defined as $\text{RR} = a \left(1 - mf\right) \left(c \cdot I_{\text{local}} + \left(1 - c\cdot I_{\text{global}}\right) \right)$, where $mf$ is the masking factor indicating the propensity of virus transmission when an agent wears a mask ($mf=0$ means 0 probability of virus spread), $c$ a constant that represents how much an agent values infections in its neighborhood, and $I_{\text{local}}$ and $I_{\text{global}}$ the proportion of infections in agent's neighborhood and in the whole network respectively.
\end{itemize}

By tuning the relative weights on these utility factors, we are able to elicit varying motivational drivers that interact to produce emergent mask-wearing behaviors. The agents learn probabilistic mask-wearing policies to maximize their utility over time using the rewards from their decisions in the unfolding pandemic environment.



\section{Results}
We analyze emerging outcomes under different configurations of the conformity, discomfort, and risk reduction weights composing the mask-wearing utility function. Experiments compare two underlying social network topologies over which the disease simulation occurs. For each parameter combination and network, simulations are initialized identically and run until conclusion of the pandemic wave. We decided to allow agents to change their decisions every 7 days, as humans do not change their masking behavior too often.

\subsection{Modeling behavior}


\begin{figure}[ht]
    \centering
\includegraphics[scale=0.45]{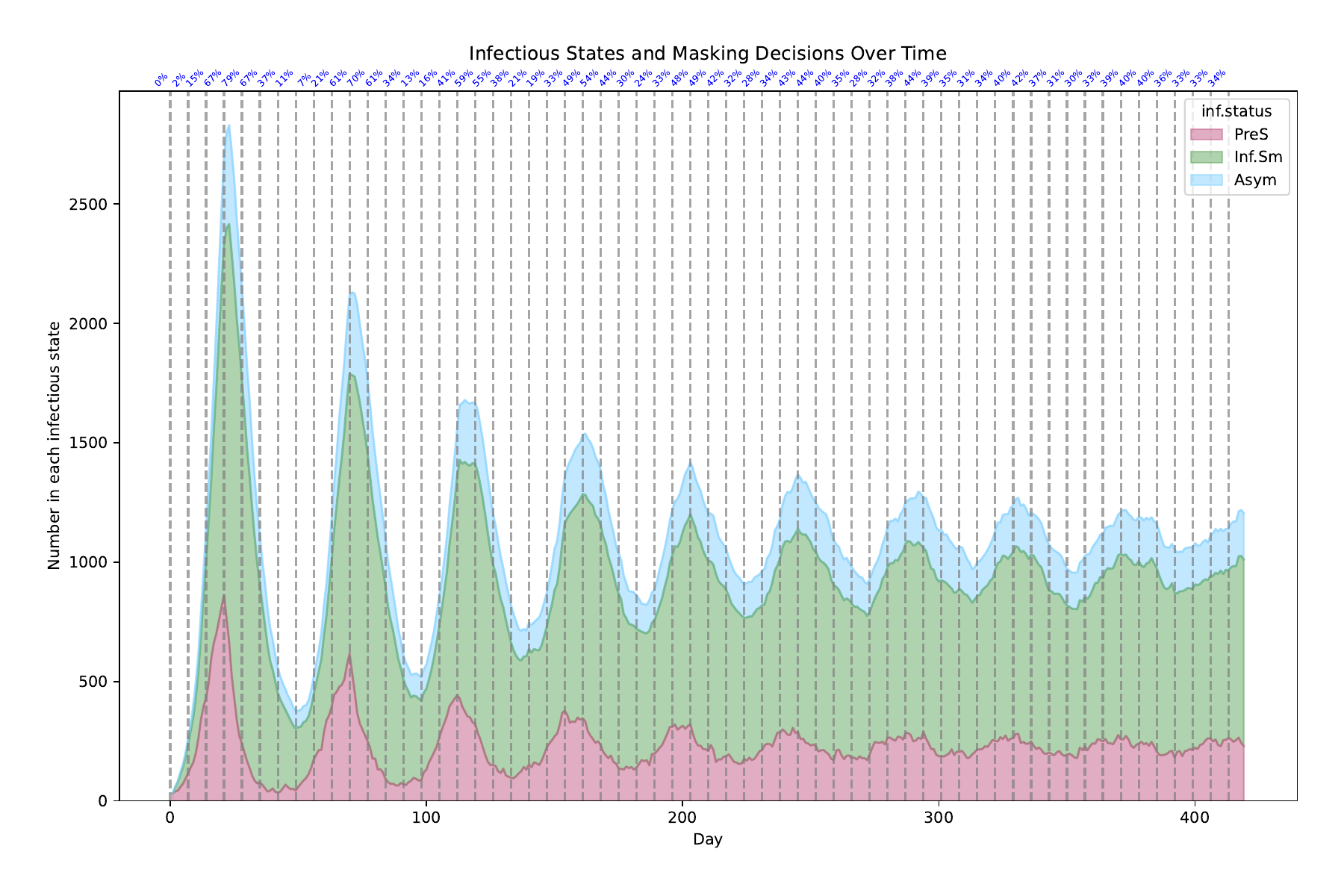}
    \caption{Epidemic evolution in the Portland network with the masking behavioral model. Local infection parameter was set to $c=0.8$, $w_1=w_2=0.5$ and $w_3=7.5$. Dashed lines represent decision making points for the agents, and in between there is a 7 day simulation period.}
    \label{fig:portland_evolution_normal}
\end{figure}

\begin{figure}[ht]
    \centering
\includegraphics[scale=0.45]{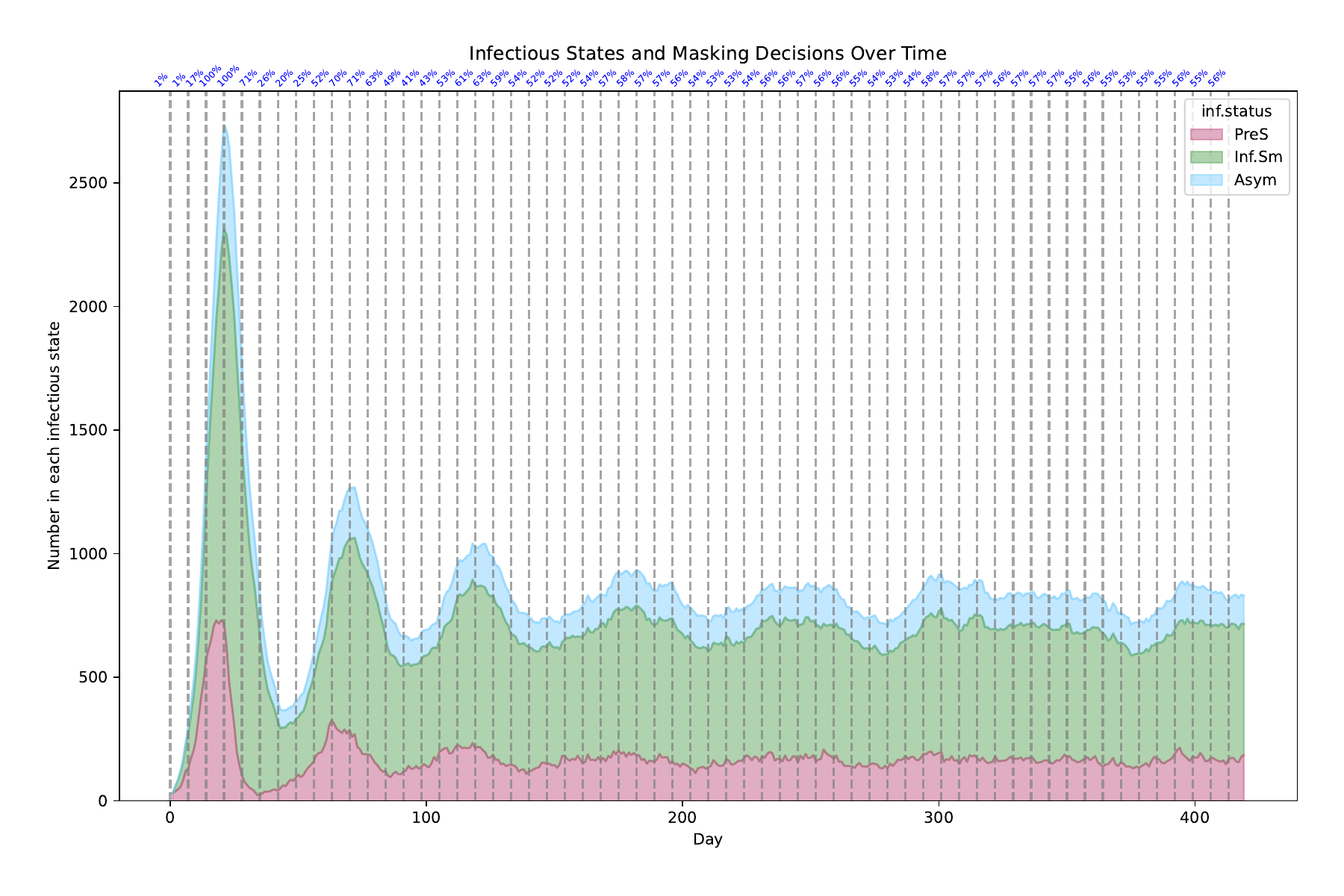}
    \caption{Epidemic evolution in the Portland network with the masking behavioral model. Local infection parameter was set to $c=0.0$, $w_1=0.5$, $w_2=0.0$ and $w_3=7.5$. Dashed lines represent decision making points for the agents, and in between there is a 7 day simulation period.}
    \label{fig:portland_evolution_global}
\end{figure}

Figure \ref{fig:portland_evolution_normal} shows the evolution of the pandemic in our base case scenario: in which we have all elements of the utility function in the Portland network. The area plot shows the number of nodes in infectious states over time and the proportion masking at each decision-step is shown along the top of the pot. In this scenario, the waves of infection are damped over time, eventually stabilizing into an equilibrium state where approximately 1 in 15 individuals are infectious on any given day. In contrast, Figure \ref{fig:portland_evolution_global} shows the evolution of the pandemic when we remove the ability for individuals to react to the local number of infections. In this scenario, masking behavior is largely coordinated across the entire network, switching rapidly between widespread masking and almost no masking, and the pandemic waves are no longer damped. Enabling individuals to observe local infections allows them to make decisions based on the risks they face. Since these vary across space, different portions of the network mask, changing the speed at which the pandemic can spread and forcing the pandemic waves to fall out of sync in these areas, damping the overall number of cases. This provides one way in which individuals, reacting to local conditions can 'flatten the curve' and lower the peak burden on emergency services.

\subsection{Masking Assortativity}

\begin{figure}[ht]
    \centering
    \begin{minipage}{0.45\textwidth}
        \centering
        \includegraphics[scale=0.3]{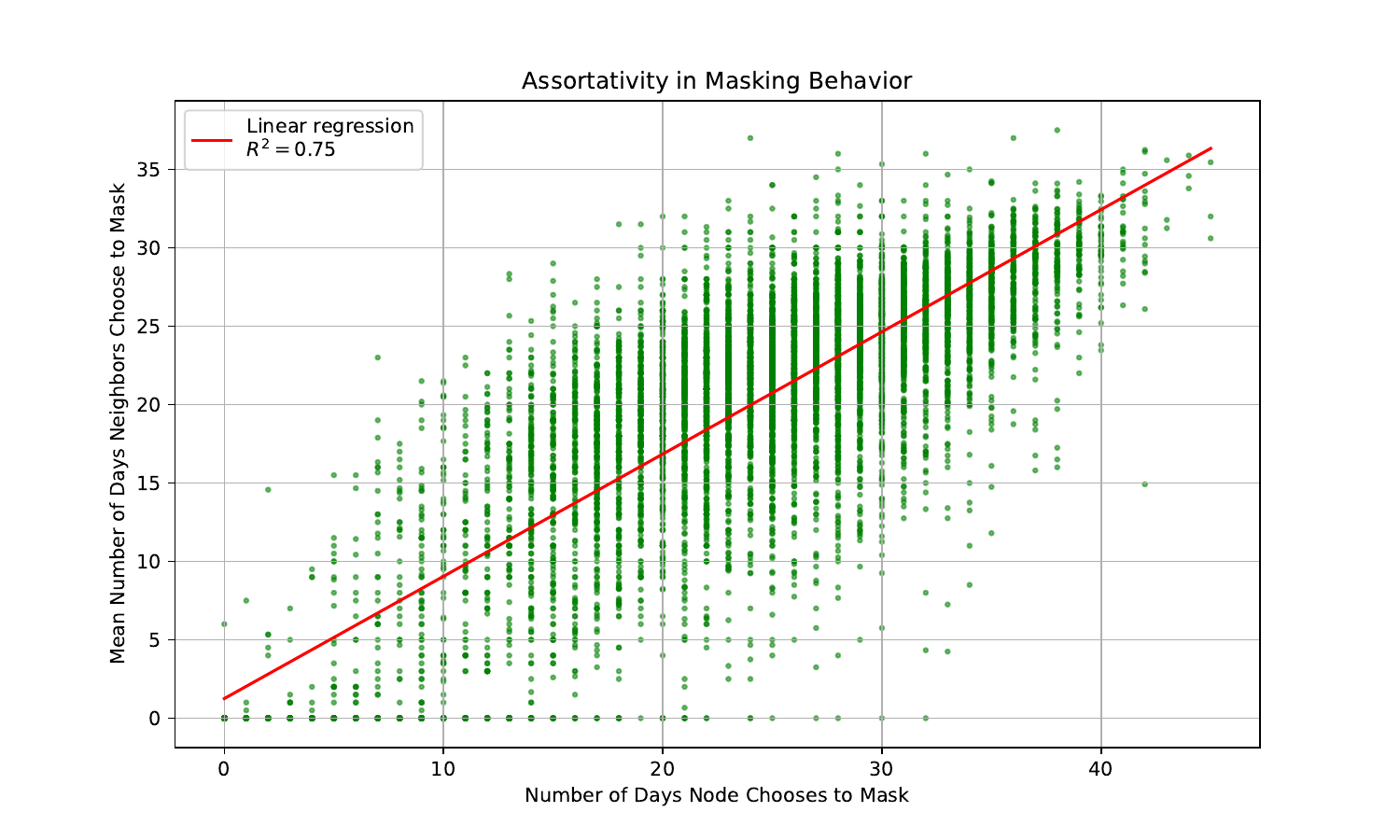}
        \caption{Assortativity in the Porltand network with local infection parameter $c=0.8$, $w_1=w_2=0.5$ and $w_3=7.5$.}
        \label{fig:assortativity_normal}
    \end{minipage}\hfill
    \begin{minipage}{0.45\textwidth}
        \centering
        \includegraphics[scale=0.3]{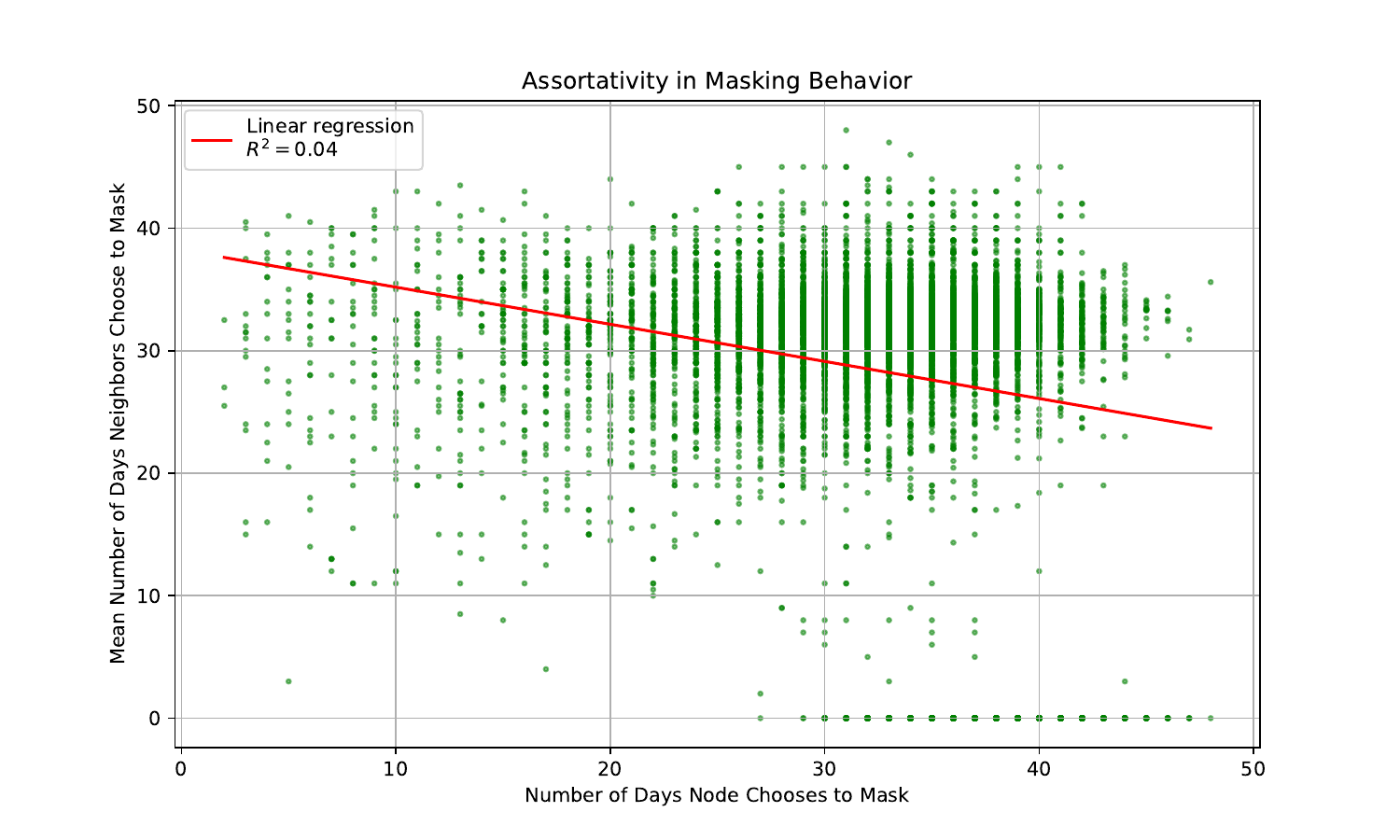}
        \caption{Assortativity in the Porltand network with local infection parameter $c=0.0$, $w_1=0.5$, similarity parameter $w_2=0.0$ and $w_3=7.5$.}
        \label{fig:assortativity_global}
    \end{minipage}
\end{figure}


Figure \ref{fig:assortativity_normal} shows masking assortativity plots using the Portland network for two conditions: the base case in which individuals have access to local and global information and a scenario where they can only observe the global state. These plots show how masking behavior of a node's neighbors changes as a function of that node's behavior across the entire duration of the simulation. The upward sloping line for the local information condition shows that masking is assortative: that masking behavior clusters together with some regions of the network masking and other regions not masking. The gradient of the line is 0.78: for each day spent masking, on average every neighbor will spend 0.78 days masking. The Pearson correlation coefficient is 0.87, indicating that vast majority of the variation in individual masking behavior is captured by the behavior of neighbors (and vice-versa). In contrast, under the global only condition (Figure \ref{fig:assortativity_global}), there is weak disassortativity, with gradient of -0.30 and a Pearson correlation of -0.21. 

Coordination of masking behavior is real-world phenomena: some communities have high levels of masking while others have low levels of masking, even when facing similar pandemic conditions. There was large variation in masking adoption across US states, and people rural areas tended to wear fewer masks than those in urban areas \cite{callaghan2021rural}. Differences in the adoption of preventative measures can potentially lead to differences in outcomes: such as the high case rates observed in rural areas (relative to urban areas) \cite{zhu2023covid}. Agent-based network approaches like the one we use in this paper are able to capture these local variations, whereas population based approaches like system dynamic models using differential equations, cannot.


\section{Discussion}

We utilized an ABM that integrates disease transmission dynamics with an adaptive behavioral model of mask-wearing using several different utility functions and two network topologies: a Portland empirical network and a scale-free Barabasi-Albert network. We varied adaptive mask-wearing behaviors in our simulations using these networks and made several key observations. First, when individuals have access to local information on infections (i.e. what proportions of their neighbors are infected), the variation in preventive actions across the network caused the disease to spread at different speeds in different parts of the network, effectively damping oscillations in the number of cases. In contrast, when individuals can only react to global infection rates, there is no mechanism to damp case oscillations and they continue unabated, potentially stretching resource capacity. Second, we shows that simple utility functions can create assortative behaviors which match real-world observations.

The use of the cognitive architecture provides multiple advantages for epidemiological modeling over conventional reinforcement learning, stemming from its adaptability, minimal training needs, cognitive interpretability, computational efficiency, and flexibility. Specifically, the instance-based approach rapidly adapts predictions as new pandemic data emerge without extensive offline dataset training, enabling responsiveness to real-time shifts. Grounding learning in ACT-R principles of cognition also facilitates interpreting model mechanisms and simulated behaviors in terms of underlying psychological theory. Further, the formulation permits straightforward scaling to thousands of socially-interacting yet autonomous agents, capturing multi-faceted community phenomena like shared identity formation and conformity pressures during crises. This provides a pathway for approximating complex motivational tension between individuality and group-aligned dynamics that require coherent context-sensitive policy insights. Together, these capabilities underscore the suitability of cognitively-inspired architectures for supporting interpretable and scalable simulations of human decision processes across social, behavioral, and epidemiological domains.

Our paper is among the first to explores how adaptive mask-wearing behavior and social networks shape the dynamics of a pandemic like COVID-19, and there are several limitations. However, our approach does have limitations that signal areas for future exploration. First, we only explore mask wearing behavior. Future models could explore how short-term masking decisions impact longer-term investments like vaccination, or social-distancing. Second, we rely on on synthetic networks, which might not capture all the structural features relevant to COVID-19. Further work could look at cases where the percolation of ideas (i.e. mask-wearing) and disease percolation occur on different networks, or integrate real-world survey data construct realistic networks. Third, we do not allow for variation in risk perception and utility functions between individuals or over time. Future work could allow for variation in risk perceptions which are transmitted across contacts, or which are intrinsic to the individual, such as fatigue in complying with preventative measures. Finally, we do not calibrate our model to real-world data, limiting the applicability of our findings to policy.

Our findings highlight the complex interactions between behavioral models, learning processes and epidemic dynamics. Further analysis is required to fully understand these behavioral models, how they can be calibrated to match real-world data, and how they might be used to guide interventions in future pandemics.




\subsubsection*{Acknowledgements}
We wish to thank the National Institute of Allergies and Infectious Diseases (R01AI118705 \& R01AI160240) for providing support in projects that led to preliminary work and ideas that motivated this project. We wish to thank Ms. 
Sarah Karr, Mr. Dulani Woods and Dr. Pedro Nascimento de Lima for their assistance in conceptualizing and developing the network-based disease transmission model of our ABM.

\bibliographystyle{apalike}
\bibliography{biblio,COVID-19}

\end{document}